\documentclass[conference]{IEEEtran}
\usepackage{graphicx}
\usepackage[utf8]{inputenc}
\usepackage{listings}
\usepackage{xspace}
\usepackage{xcolor}
\usepackage{subcaption}
\usepackage{balance}
\usepackage[printwatermark]{xwatermark}

\usepackage[
bookmarks=false,
breaklinks=true,
colorlinks=true,
linkcolor=black,
citecolor=black,
urlcolor=black,
pdfpagelayout=SinglePage,
pdfstartview=Fit
]{hyperref}
\usepackage[all]{hypcap}

\title{Instant restore after a media failure}

\author{
    \IEEEauthorblockN{Caetano Sauer}
    \IEEEauthorblockA{TU Kaiserslautern, Germany\\
    {csauer@cs.uni-kl.de}}
\and
    \IEEEauthorblockN{Goetz Graefe}
    \IEEEauthorblockA{Google, Madison, WI, USA\\
    {goetzg@google.com}}
\and
    \IEEEauthorblockN{Theo H\"arder}
    \IEEEauthorblockA{TU Kaiserslautern, Germany\\
    {haerder@cs.uni-kl.de}}
}

\begin{document}

\maketitle

\begin{abstract}
Media failures usually leave database systems unavailable for several hours until recovery is complete, especially in applications with large devices and high transaction volume.
Previous work introduced a technique called single-pass restore, which increases restore bandwidth and thus substantially decreases time to repair.
Instant restore goes further as it permits read/write access to any data on a device undergoing restore---even data not yet restored---by restoring individual data segments on demand.
Thus, the restore process is guided primarily by the needs of applications, and the observed mean time to repair is effectively reduced from several hours to a few seconds.

This paper presents an implementation and evaluation of instant restore.
The technique is incrementally implemented on a system starting with the traditional ARIES design for logging and recovery.
Experiments show that the transaction latency perceived after a media failure can be cut down to less than a second and that the overhead imposed by the technique on normal processing is minimal.
The net effect is that a few “nines” of availability are added to the system using simple and low-overhead software techniques.

\end{abstract}



\section{Introduction}\label{sec:Introduction}
Advancements in hardware technology have significantly improved the performance of database systems over the last decade, allowing for throughput in the order of thousands of transactions per second and data volumes in the order of petabytes.
Availability, on the other hand, has not seen drastic improvements, and the research goal postulated by Jim Gray in his ACM Turing Award Lecture of a system “unavailable for less than one second per hundred years” \cite{bib:GrayTuringLecture} remains an open challenge.
Improvements in reliable hardware and data center technology have contributed significantly to the availability goal, but proper software techniques are required to not only avoid failures but also repair failed systems as quickly as possible.
This is especially relevant given that a significant share of failures is caused by human errors and unpredictable defects in software and firmware, which are immune to hardware improvements \cite{bib:Heisenbugs,bib:ROC}.
In the context of database logging and recovery, the state of the art has unfortunately not changed much since the early 90’s, and no significant advancements were achieved in the software front towards the availability goal.

Instant restore is a technique for media recovery that drastically reduces mean time to repair by means of simple software techniques.
It works by extending the write-ahead logging mechanism of ARIES \cite{bib:Mohan92ARIES} and, as such, can be incrementally implemented on the vast majority of existing database systems.
The key idea is to introduce a different organization of the log archive to enable efficient on-demand, incremental recovery of individual data pages.
This allows transactions to access recovered data from a failed device orders of magnitude faster than state-of-the-art techniques, all of which require complete restoration of the entire device before access to the application's working set is allowed.

\begin{figure}
\centering
\includegraphics[width=\linewidth]{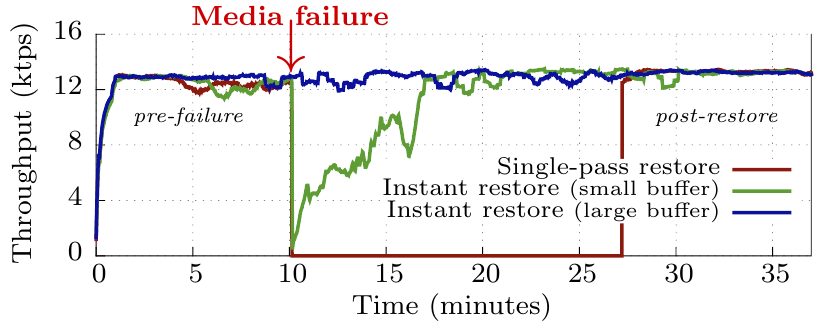}
  \caption{Effect of instant restore}
\label{fig:tput_simple}
\vspace{-0.3cm}
\end{figure}

The problem of inefficient media recovery in state-of-the-art techniques, including ARIES and its optimizations, can be attributed to two major deficiencies.
First, the media recovery process has a very inefficient random access pattern, which in practice encourages excessive redundancy and high-frequency incremental backups---solutions that only alleviate the problem instead of eliminating it.
The second deficiency is that the recovery process is not incremental and requires full recovery before any data can be accessed---on-demand schedules are not possible and there is no prioritization scheme to make most needed data available earlier.
Previous work addressed the first problem with a technique called single-pass restore \cite{bib:SinglePassRestore}, while the present paper focuses on the second one.


The effect of instant restore is illustrated in Figure \ref{fig:tput_simple}, where transaction throughput is plotted over time and a media failure occurs after 10 minutes.
In single-pass restore, as in ARIES, transaction processing halts until the device is fully restored (the red line in the chart), while instant restore continues processing transactions, using them to guide the restore process (blue and green lines).
In a scenario where the application working set fits in the buffer pool (blue line), there is actually no visible effect on transaction throughput.
We emphasize that traditional ARIES media recovery would take much longer than the scale used in the diagram; therefore, the baseline used to measure our present work is single-pass restore.
More detailed and comprehensive experiments are presented in Section~\ref{sec:Experiments}.

In the remainder of this paper, Section~\ref{sec:RelatedWork} describes related work, both previous work leading to the current design as well as competing approaches.
Then, Section~\ref{sec:Restore} describes the instant restore technique.
Finally, Section~\ref{sec:Experiments} presents an empirical evaluation, while Section~\ref{sec:Conclusion} concludes this paper. 

A high-level description of instant restore was previously published in a book chapter \cite{bib:InstantRecoverySynthesis2nd} among related instant recovery techniques.
The additional contribution here is a much more detailed discussion of the design---including practical implementation aspects---as well as the first empirical evaluation of the technique with an open-source prototype.

\section{Related work}\label{sec:RelatedWork}
This section starts by establishing the scope of our work with respect to failure classes considered in transaction recovery literature and defining basic assumptions.
Afterwards, we discuss existing media recovery techniques, focusing mainly on the limitations that will be addressed later in Section~\ref{sec:Restore}.

\renewcommand{\arraystretch}{1.3}
\begin{table}[]
\centering
\caption{Failure classes, their causes, and effects}
\label{tab:FailureClasses}
\begin{tabular}{|l|p{0.30\linewidth}|p{0.20\linewidth}|l|}
\hline
\textbf{Failure class} & \textbf{Loss}                    & \textbf{Typical cause}      & \textbf{Response} \\ \hline
Transaction            & Single-transaction progress      & Deadlock                    & Rollback          \\ \hline
System                 & Server process (in-memory state) & Software fault, power loss & Restart           \\ \hline
Media                  & Stored data                      & Hardware fault              & Restore           \\ \hline
Single page            & Local integrity                  & Partial writes, wear-out    & Repair            \\ \hline
\end{tabular}
\end{table}

\subsection{Failure classes and assumptions}\label{sec:FailureClasses}
Database literature traditionally considers three classes of database failures \cite{bib:HaerderReuter83}, which are summarized in Table~\ref{tab:FailureClasses} (along with single-page failures, a fourth class to be discussed in Section~\ref{sec:SinglePageRepair}).
In the scope of this paper, it is important to distinguish between system and media failures, which are conceptually quite different in their causes, effects, and recovery measures.
System failures are usually caused by a software fault or power loss, and what is lost---hence what must be recovered---is the state of the server process in main memory; this typically entails recovering page images in the buffer pool (i.e., ``repeating history'' \cite{bib:Mohan92ARIES}) as well as lists of active transactions and their acquired locks, so that they can be properly aborted.
The process of recovering from system failures is called \emph{restart}.

Instant restart \cite{bib:InstantRecoverySynthesis2nd} is an orthogonal technique that provides on-demand, incremental data access following a system failure.
While the goals are similar, the design and implementation of instant restore require quite different techniques.

In a media failure, which is the focus here, a persistent storage device fails but the system might continue running, serving transactions that only touch data in the buffer pool or on other healthy devices.
If the system and media failures happen simultaneously, or perhaps one as a cause of the other, their recovery processes are executed independently, and, by recovering pages in the buffer pool, the processes coordinate transparently.
Readers are referred to the literature for further details \cite{bib:WeikumVossen}.

The present work makes the same assumptions as most prior research on database recovery.
The log and its archival copy reside on ``stable storage'', i.e., they are assumed to never fail.
We consider media failure on the database device only, i.e., the permanent storage location of data pages.
Recovery from such failures requires a backup copy (possibly days or weeks old) of the lost device and all log records since the backup was taken; such log records may reside either in the active transaction log or in the log archive.
The process of recovering from media failures is called \emph{restore}.
The following sections briefly describe previous restore methods.

\subsection{ARIES restore}\label{sec:ARIESrestore}
\begin{figure}
\centering
\includegraphics[width=\linewidth]{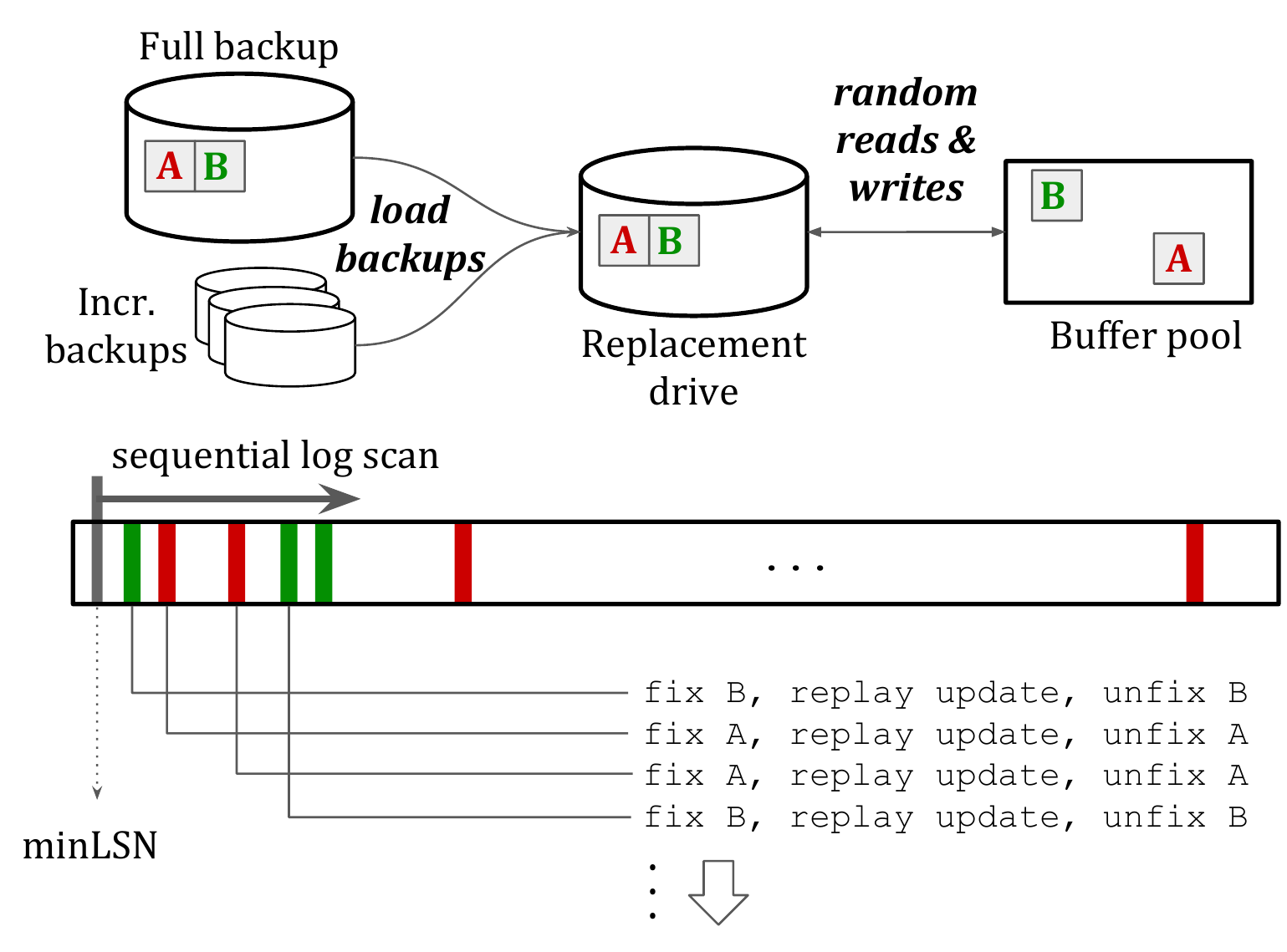}
  \caption{Random access pattern of ARIES restore}
\label{fig:aries_restore}
\vspace{-0.3cm}
\end{figure}

Techniques to recover databases from media failures were initially presented in the seminal work of Gray~\cite{bib:JimGrayNotes} and later incorporated into the ARIES family of recovery algorithms \cite{bib:Mohan92ARIES}.
In ARIES, restore after a media failure first loads a backup image and then applies a redo log scan, similar to the redo scan of restart after a system failure.
Fig.~\ref{fig:aries_restore} illustrates the process, which we now briefly describe.
After loading full and incremental backups into the replacement device, a sequential scan is performed on the log archive and each update is replayed on its corresponding page in the buffer pool.
A global \emph{minLSN} value (called ``media recovery redo point'' by Mohan et al.~\cite{bib:Mohan92ARIES}) is maintained on backup devices to determine the begin point of the log scan.

Because log records are ordered strictly by LSN, pages are read into the buffer pool in random order, as illustrated in the restoration of pages A and B in Fig.~\ref{fig:aries_restore}.
Furthermore, as the buffer pool fills up, they are also written in random order into the replacement device, except perhaps for some minor degree of clustering.
As the log scan progresses, evicted pages might be read multiple times, also randomly.
This mechanism is quite inefficient, especially for magnetic drives with high access latencies.
Thus, it is no surprise that multiple hours of downtime are required in systems with high-capacity drives and high transaction rates \cite{bib:SinglePassRestore}.

Another fundamental limitation of the ARIES restore algorithm is that it is not incremental, i.e., pages cannot be restored to their most up-to-date version one-by-one and made available to running transactions incrementally.
As shown in the example of Fig.~\ref{fig:aries_restore}, the last update to page A may be at the very end of the log; thus, page A will be out-of-date until almost the end of the long log scan.
Some optimizations may alleviate this situation (e.g., reusing checkpoint information), but there is no general mechanism for incremental restoration.
Furthermore, even if pages could somehow be released incrementally when their last update is replayed, the hottest pages of the application working set are most likely to be released only at the very end of the log scan, and probably not even then, because they might contain updates of uncommitted transactions and thus require subsequent undo.
This leads to yet another limitation of this log-scan-based approach: even if pages could be restored incrementally, there is no effective way to provide on-demand restoration, i.e., to restore most important pages first.

Despite a variety of optimizations proposed to the basic ARIES algorithm \cite{bib:Mohan92ARIES,bib:Mohan93Archiving,bib:MohanRemoteBackup}, none of them solves these problems in a general and effective manner.
In summary, all proposed techniques that enable earlier access to recovered data items suffer from the same problem: early access is only provided for data for which early access is not really needed---hot data in the application working set is not prioritized and most accesses must wait for complete recovery.

Finally, industrial database systems that implement ARIES recovery suffer from the same problems.
IBM’s DB2 speeds up log replay by sorting log records after restoring the backup and before applying the log records to the replacement database \cite{bib:IBMFastLogApply}.
While a sorted log enables a more efficient access pattern, incremental and on-demand restoration is not provided.
Furthermore, the delay imposed by the offline sort may be as high as the total downtime incurred by the traditional method.

Oracle attempts to eliminate the overhead of reading incremental backups by incrementally maintaining a full backup image \cite{bib:OracleBackups}.
While this makes the access pattern slightly more efficient, it does not address the deficiencies discussed earlier.


\subsection{Replication}
Given the extremely high cost of media recovery in existing systems, replication solutions such as disk mirroring or RAID \cite{bib:GrayDiskShadowing,bib:RAIDCSUR} are usually employed in practice to increase mean time to failure.
However, it is important to emphasize that, from the database system's perspective, a failed disk in a redundant array does not constitute a media failure as long as it can be repaired automatically.
Restore techniques aim to improve mean time to repair whenever a failure that cannot be masked by lower levels of the system occurs.
Therefore, replication techniques can be seen largely as orthogonal to media restore techniques as implemented in database recovery mechanisms.

However, a substantial reduction in mean time to repair, especially if done solely with simple software techniques, opens many opportunities to manage the trade-off between operational costs and availability.
One option can be to maintain a highly-available infrastructure (with whatever costs it already requires) while availability is increased by deploying software with more efficient recovery.
Alternatively, replication costs can be reduced (e.g., downgrading RAID-10 into RAID-5) while maintaining the same availability.
Such level of flexibility, with solutions tackling both mean time to failure and mean time to repair, are essential in the pursuit of Gray's availability goal \cite{bib:GrayTuringLecture}, especially considering the impact of human errors and unpredictable failures that occur in large deployments \cite{bib:Heisenbugs,bib:ROC,bib:ReliabilityFreshLook}.


Early work on in-memory databases focused mainly on restart after a system failure, employing traditional backup and log-replay techniques for media recovery \cite{bib:EichMMDBRecoverySurvey,bib:LehmanCareyMMDBRecovery}.
The work of Levi and Silberschatz \cite{bib:LevySilberschatzIncrRecovery} was among the first to consider the challenge of incremental restart after a system failure.
While an extension of their work for media recovery is conceivable, it would not address the efficiency problem discussed in Section~\ref{sec:Introduction}.
Thus, it would, in the best case and with a more complex algorithm, perform no better than the related work discussed later in Section~\ref{sec:SinglePageRepair}.

\subsection{In-memory databases}
Recent proposals for recovery on both volatile and non-volatile in-memory systems usually ignore the problem of media failures, employing the unspecific term ``recovery'' to describe system restart only \cite{bib:HekatonDEBU,bib:RethinkingOLTPRecovery,bib:HyPer,bib:LetsTalk,bib:SOFORT}.
Therefore, recovery from media failures in modern systems either relies on the traditional techniques or is simply not supported, employing replication as the only means to maintain service upon storage hardware faults.
As discussed above, while relying on replication is a valid solution to increase mean time to failure, a highly available system must also provide efficient repair facilities.
In this aspect, traditional database system designs---using ARIES physiological logging and buffer management---provide more reliable behavior.
Therefore, we believe that improving traditional techniques for more efficient recovery with low overhead on memory-optimized workloads is an important open research challenge.

\subsection{Single-page repair}\label{sec:SinglePageRepair}

\begin{figure}
\centering
\includegraphics[width=\linewidth]{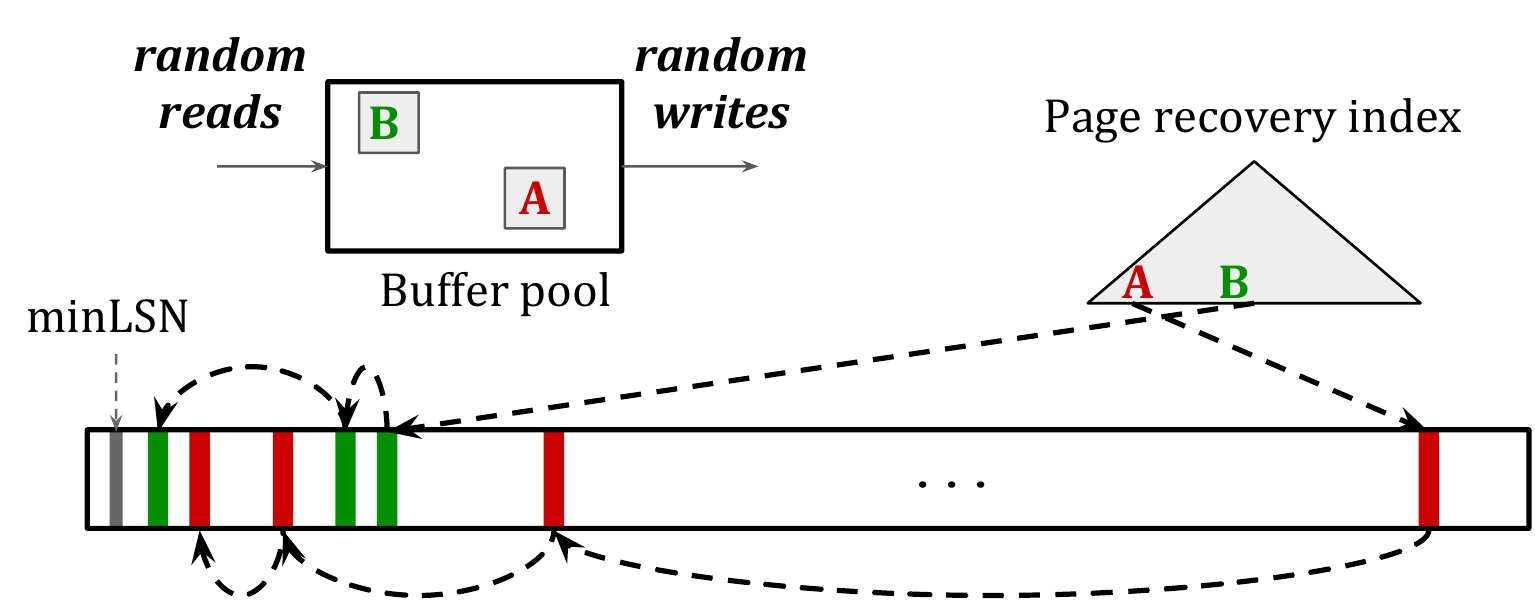}
  \caption{Single-page repair}
\label{fig:single_page_rec}
\vspace{-0.3cm}
\end{figure}

Single-page failures are considered a fourth class of database failures \cite{bib:SinglePageRecovery}, along with the other classes summarized in Table~\ref{tab:FailureClasses}.
It covers failures restricted to a small set of individual pages of a storage device and applies online localized recovery to that individual page instead of invoking media recovery on the whole device.
The single-page repair algorithm, illustrated in Fig.~\ref{fig:single_page_rec} (with backup and replacement devices omitted for simplification), has two basic requirements: first, the LSN of the most recent update of each page is known (i.e., the current PageLSN value) without having to access the page; second, starting from the most recent log record, the complete history of updates to a page can be retrieved.
The former requirement can be provided with a page recovery index---a data structure mapping page identifiers to their most recent PageLSN value.
Alternatively, the current PageLSN can be stored together with the parent-to-child node pointer in a B-tree data structure \cite{bib:SelfHealingIndexes}.
The latter requirement is provided by per-page log record chains, which are straight-forward to maintain using the PageLSN fields in the buffer pool.
For each page update, the LSN of the last log record to affect the same page (i.e., the pre-update PageLSN value) is recorded in the log record; this allows the history of updates do be derived by following the resulting chain of backward pointers.
We refer to the paper for further details~\cite{bib:SinglePageRecovery}.

In principle, single-page repair could be used to recover from a media failure, by simply repairing each page of the failed device individually.
One advantage of this technique is that it yields incremental and on-demand restore, addressing the second deficiency of traditional media recovery algorithms mentioned in Section~\ref{sec:Introduction}.
To illustrate how this would work in practice, consider the example of Fig.~\ref{fig:single_page_rec}.
If the first page to be accessed after the failure is A, it would be the first to be restored.
Using information from the page recovery index (which can be maintained in main memory or fetched directly from backups), the last red log record on the right side of the diagram would be fetched first.
Then, following the per-page chain, all red log records until \emph{minLSN} would be retrieved and replayed in the backup image of page A, thus yielding its most recent version to running transactions.

While the benefit of on-demand and incremental restore is a major advantage over traditional ARIES recovery, this algorithm still suffers from the first deficiency discussed in Section~\ref{sec:Introduction}---namely the inefficient access pattern.
The authors of the original publication even foresee the application to media failures \cite{bib:SinglePageRecovery}, arguing that while a page is the unit of recovery, multiple pages can be repaired in bulk in a coordinated fashion.
However, the access pattern with larger restoration granules would approach that of traditional ARIES restore---i.e., random access during log replay.
Thus, while the technique introduces a very useful degree of flexibility, it does not provide a unified solution for the two deficiencies discussed.

\subsection{Single-pass restore}

\begin{figure}
\centering
\includegraphics[width=\linewidth]{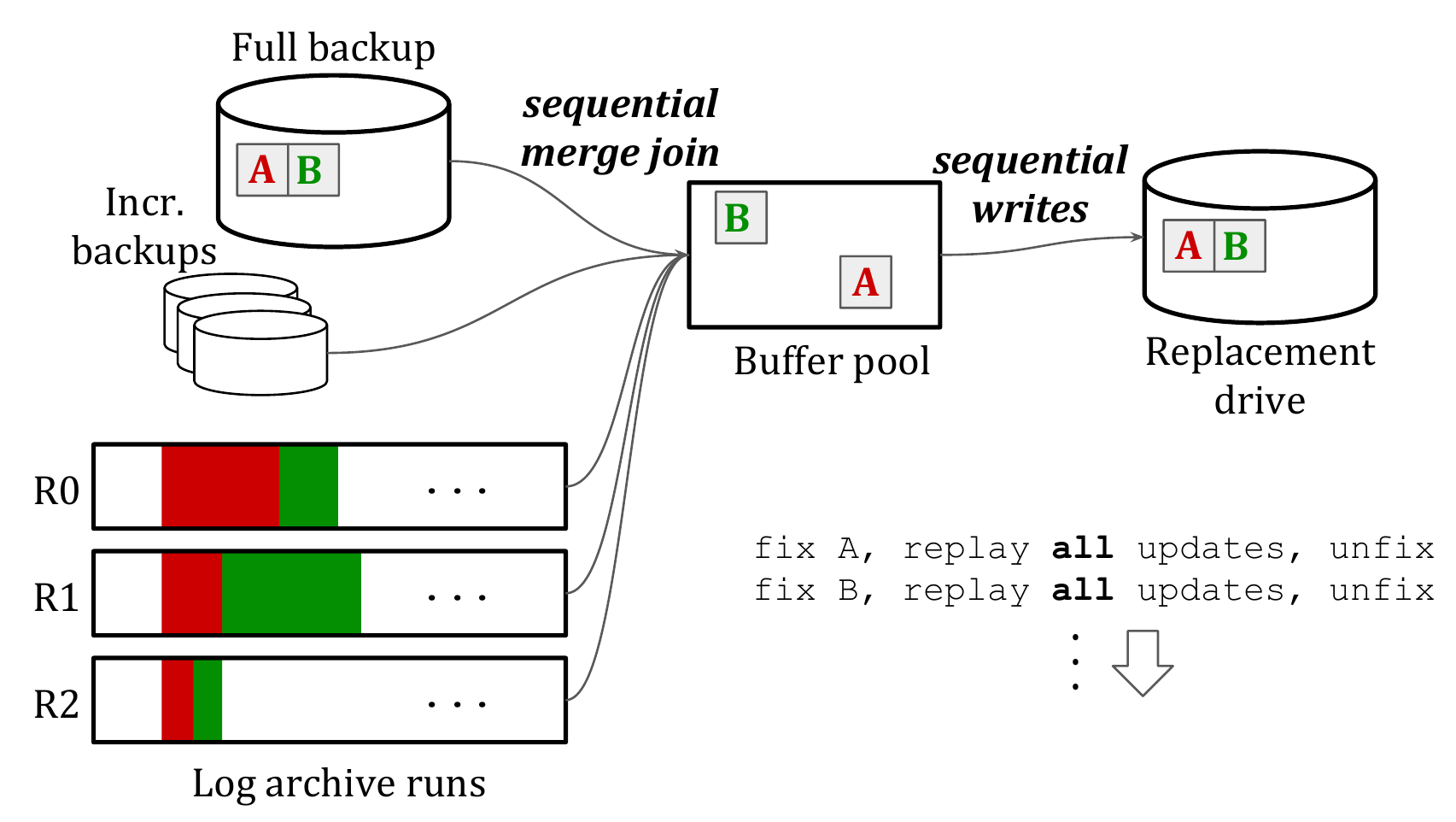}
  \caption{Single-pass restore}
\label{fig:single_pass_rest}
\vspace{-0.3cm}
\end{figure}

Our previous work introduced a technique called single-pass restore, which aims to perform media recovery in a single sequential pass over both backup and log archive devices \cite{bib:SinglePassRestore}.
Eliminating random access effectively addresses the first deficiency discussed in Section~\ref{sec:Introduction}.
This is achieved by partially sorting the log on page identifiers, using a stable sort to maintain LSN order within log records of the same page.
The access pattern is essentially the same as that of a sort-merge join: external sort with run generation and merge followed by another merge between the two inputs---log and backup in the media recovery case.

The idea itself is as old as the first recovery algorithms (see Section~5.8.5.1 of Gray's paper~\cite{bib:JimGrayNotes}) and is even employed in DB2's ``fast log apply'' \cite{bib:IBMFastLogApply}.
However, the key advantage of single-pass restore is that the two phases of the sorting process---run generation and merge---are performed independently: sorted runs are generated during the log archiving process (i.e., moving log records from the latency-optimized transaction log device into high-capacity, bandwidth-optimized secondary storage) with negligible overhead; the merge phase, on the other hand, happens both asynchronously as a maintenance service and also during media recovery, in order to obtain a single sorted log stream for recovery.
Importantly, merging runs of the log archive and applying the log records to backed-up pages can be done in a single sequential pass, similar to a merge join.
The process is illustrated in Fig.~\ref{fig:single_pass_rest}.
We refer to the original publication for further details \cite{bib:SinglePassRestore}.

Having addressed the access pattern deficiency of media recovery algorithms, single-pass restore still leaves open the problem of incremental and on-demand restoration.
Nevertheless, given its superiority over traditional ARIES restore (see \cite{bib:SinglePassRestore} and \cite{bib:InstantRecoverySynthesis2nd} for an in-depth discussion), it is a promising approach to use as starting point in addressing the two deficiencies in a unified way.
Therefore, as mentioned in Section~\ref{sec:Introduction}, single-pass restore is taken as the baseline for the present work.

\subsection{Summary of related work}
As the previous sections discussed, none of the state-of-the-art media recovery schemes is able to effectively eliminate the two main deficiencies of traditional ARIES: the inefficient access pattern and the lack of early access to important data before complete recovery.
Ideally, a restore mechanism would combine the incremental availability and on-demand schedule provided by single-page repair with the efficient, bulk access pattern of single-pass restore.
Moreover, this combination should allow for a continuous adjustment between these two behaviors and a simple adaptive technique should make the best decision dynamically based on system behavior.
These challenges are addressed by instant restore, which we describe next.

\clearpage

\section{Instant restore}\label{sec:Restore}
The main goal of instant restore is to preserve the efficiency of single-pass restore while allowing more fine-granular restoration units (i.e., smaller than the whole device) that can be recovered incrementally and on demand.
We propose a generalized approach based on segments, which consist of contiguous sets of data pages.
If a segment is chosen to be as large as a whole device, our algorithm behaves exactly like single-pass restore; on the other extreme, if a segment is chosen to be a single page, the algorithm behaves like single-page repair.
As discussed in this section and evaluated empirically in Section~\ref{sec:Experiments}, the optimal restore behavior lies somewhere between these two extremes, and simple adaptive techniques are proposed to robustly deliver good restore performance without turning knobs manually.

This section is divided in four parts: first, we introduce the log data structure employed to provide efficient access to log records belonging to a given segment or page; after that, we present the restore algorithm based on this data structure; then, we discuss techniques to choose the best segment size dynamically and thus optimize restore behavior; finally, we discuss the issue of coordinating processes of different recovery modes (e.g., restart and restore at the same time) as well as concurrent threads of the same recovery mode (e.g., multiple restore threads).

\subsection{Indexed log archive}\label{sec:LogArchive}
In order to restore a given segment incrementally, instant restore requires efficient access to log records pertaining to pages in that segment.
In single-page repair, such access is provided for individual pages, using the per-page chain among log records \cite{bib:SinglePageRecovery}.
As already discussed, this is not efficient for restoration units much larger than a single page.
Therefore, we build upon the partially sorted log archive organization introduced in single-pass restore \cite{bib:SinglePassRestore}.

In instant restore, the partially sorted log archive is extended with an index.
The log archiving process sorts log records in an in-memory workspace and saves them into runs on persistent storage.
These runs must then be indexed, so that log records of a given page or segment identifier can be fetched directly.
Sorting and indexing of log records is done online and without any interference to transaction processing, in addition to standard log archiving tasks such as compression.

Fig.~\ref{fig:indexed_log_archive} illustrates indexed log archive runs and a range lookup for a segment containing pages G to K.
As explained in previous work \cite{bib:SinglePassRestore}, runs must are mapped to contiguous LSN ranges to simplify log archiving restart and garbage collection.
In an index lookup for instant restore, the set of runs to consider would be restricted by the given \emph{minLSN} (see Section~\ref{sec:ARIESrestore}) of the backup image, since runs older than that LSN are not needed.
Furthermore, bloom filters can be appended to each run to restrict this set even further.
The result of the lookup in each indexed run is then fed into a merge process that delivers a single stream of log records sorted primarily by page identifier and secondarily by LSN.
This stream can then be used by the restore algorithm to replay updates on backed-up images of segments.

Multiple choices exist for the physical data structure of the indexed log archive.
Ideally, the B-tree component of the indexing subsystem can be reused, but there is an important caveat in terms of providing atomicity and durability to this structure.
A typical index relies on write-ahead logging, but that is not an option for the indexed log archive because it would introduce a kind of self-reference loop---updates to the log data structure itself would have to be logged and used later on for recovery.
This self-reference loop could be dealt with by introducing special logging and recovery modes (e.g., a separate ``meta''-log for the indexed log archive), but the resulting algorithm would be too cumbersome.

A more viable solution is to rely on an atomic data structure, like the shadow-based B-tree proposed by Rodeh \cite{bib:BtreeShadowing}.
Since the log archive is mostly a read-only data structure, where the only writes are bulk appends or merges, such shadowing approaches are perfectly suitable.
In our prototype, we chose a different approach, where each partition of the log archive is maintained in its own read-only file; temporary shadow files are then used for merges and appends.
In this scheme, atomicity is provided by the file rename operation, which is atomic in standard filesystems \cite{bib:libcRename}.

\subsection{Restore algorithm}\label{sec:RestoreAlgorithm}
\begin{figure}
\centering
\includegraphics[width=\linewidth]{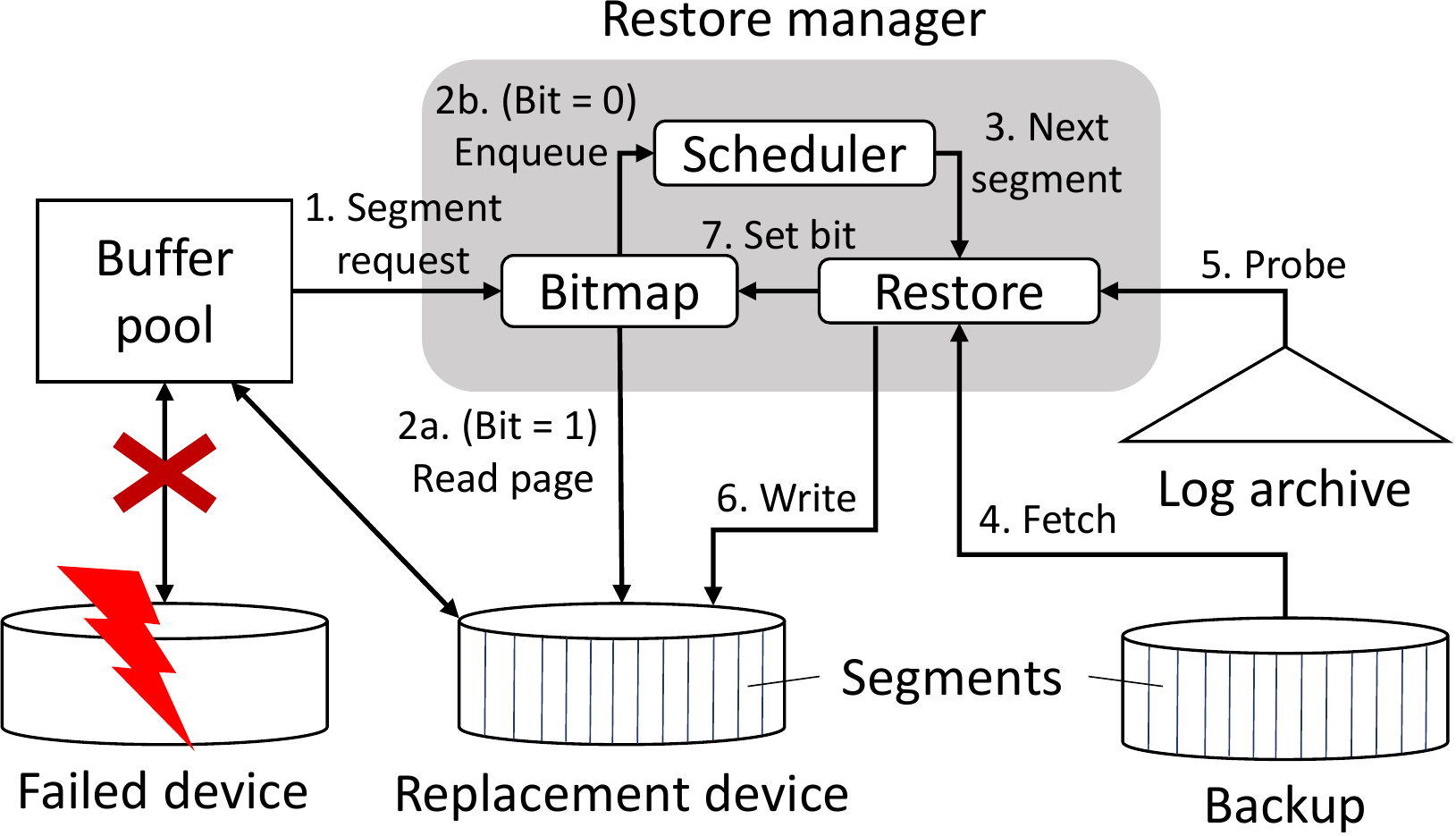}
\caption{Instant restore flow chart}
\label{fig:flowchart}
\vspace{-0.3cm}
\end{figure}

When a media failure is detected, a restore manager component is initialized and all page read and write requests from the buffer pool are intercepted by this component.
The diagram in Fig.~\ref{fig:flowchart} illustrates the interaction of the restore manager with the buffer pool and all persistent devices involved in the restore process: failed and replacement devices, log archive, and backup.
For reasons discussed in previous work \cite{bib:SinglePassRestore}, incremental backups are made obsolete by the partially sorted log archive; thus, the algorithm performs just as well with full backups only.
Nevertheless, incremental backups can be easily incorporated, and the description below considers a single full backup without loss of generality.

\begin{figure}
\centering
\includegraphics[width=\linewidth]{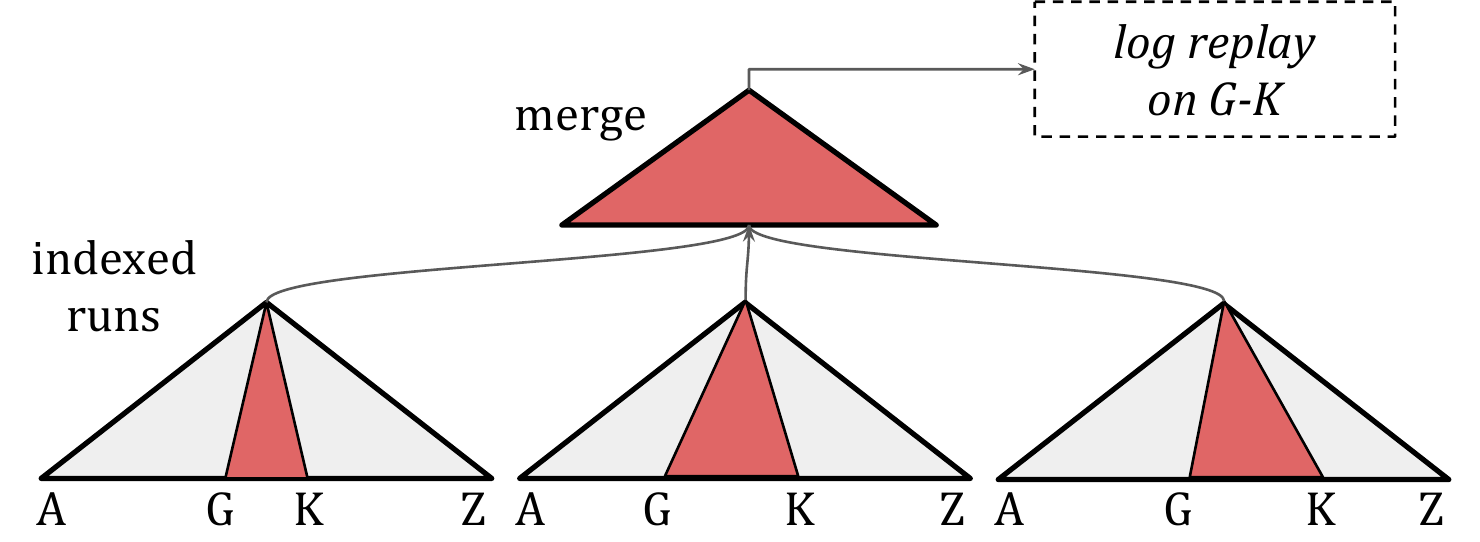}
  \caption{Indexed log archives and their use in instant restore}
\label{fig:indexed_log_archive}
\end{figure}

In the following discussion, the numbers in parentheses refer to the numbered steps in Fig.~\ref{fig:flowchart}.
The restore manager keeps track of which segments were already restored using a segment recovery bitmap, which is initialized with zeros.
When a page access occurs, the restore manager first looks up its segment in the bitmap (1).
If set to one, it indicates that the segment was already restored and can be accessed directly on the replacement device (2a).
If set to zero, a segment restore request is enqueued into a restore scheduler (2b), which coordinates the restoration of individual segments (3).

To restore a given segment, an older version is first fetched from the backup directly (4).
This is in contrast to ARIES restore, which first loads entire backups into the replacement device and then reads pages from there \cite{bib:Mohan92ARIES}.
This has the implication that backups must reside on random-access devices (i.e., not on tape) and allow direct access to individual segments, which might require an index if backup images are compressed.
These requirements, which are also present in single-page repair \cite{bib:SinglePageRecovery}, seem quite reasonable given the very low cost per byte of current high-capacity hard disks.
For moderately-sized databases, it is even advisable to maintain log archive and backups on flash storage.

While the backed-up image of a segment is loaded, the indexed log archive data structure is probed for the log records pertaining to that segment (5).
This initializes the merge logic illustrated in Fig.~\ref{fig:indexed_log_archive}.
Then, log replay is performed to bring the segment to its most recent state, after which it can be written back into a replacement device (6).

Finally, once a segment is restored, the bitmap is updated (7) and all pending read and write requests can proceed.
Typically, a requested page will remain in the buffer pool after its containing segment is restored, so that no additional I/O access is required on the replacement device.

All read and write operations described above---log archive index probe, segment fetch, and segment write after restoration---happen asynchronously with minimal coordination.
The read operations are essentially merged index scans---a very common pattern in query processing \cite{bib:GraefeQueryEvaluation}.
The write of a restored segment is also easily made asynchronous, whereby the only requirement is that marking a segment as restored on the bitmap, and consequently enabling access by waiting threads, be done by a callback function after completion of the write.

\begin{figure}
\centering
\includegraphics[width=\linewidth]{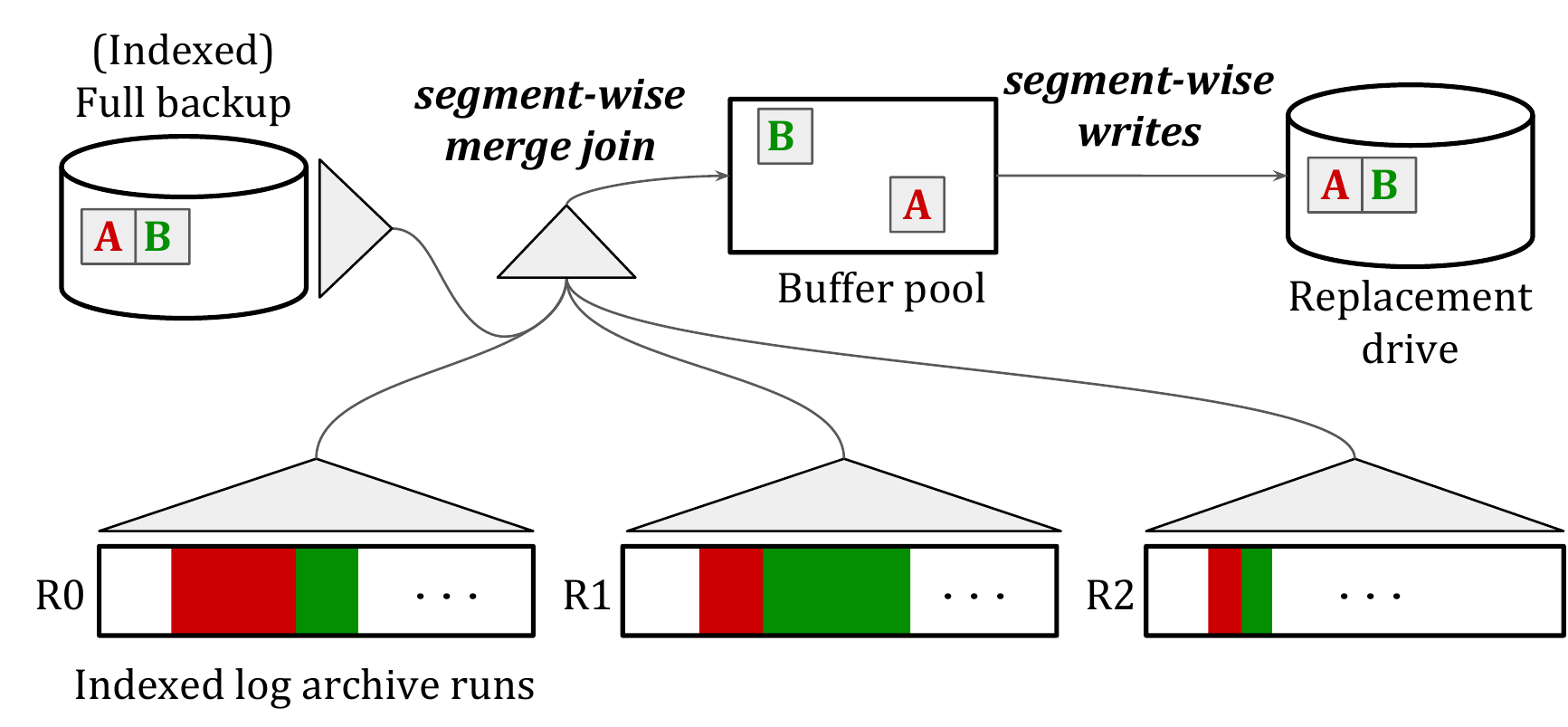}
  \caption{Instant restore}
\label{fig:instant_rest}
\vspace{-0.3cm}
\end{figure}

To illustrate the access pattern of instant restore, similarly to the diagrams in Section~\ref{sec:RelatedWork}, Fig.~\ref{fig:instant_rest} shows an example scenario with three log archive runs and two pages, A and B, belonging to the same segment.
The main difference to the previous diagrams is the segment-wise, incremental access pattern, which delivers the efficiency of pure sequential access with the responsiveness of on-demand random reads.

Using this mechanism, user transactions accessing data either in the buffer pool or on segments already restored can execute without any additional delay, whereby the media failure goes completely unnoticed.
Access to segments not yet restored are used to guide the restore process, triggering the restoration of individual segments on demand.
As such, the time to repair observed by transactions accessing data not yet restored is multiple orders of magnitude lower than the time to repair the whole device.
Furthermore, time to repair observed by an individual transaction is independent of the total capacity of the failed device.
This is in contrast to previous methods, which require longer downtime for larger devices.

\subsection{Latency vs.~bandwidth trade-off}
One major contribution of instant restore is that it generalizes single-page repair and single-pass restore, providing a continuum of choices between the two.
In order to optimize restore behavior, the restore manager must adaptively and robustly choose the best option within this continuum.
In practice, this boils down to choosing the correct granularity of access to both backup and log archive, in order to balance restore latency and bandwidth.

\begin{figure}
\centering
\includegraphics[width=0.8\linewidth]{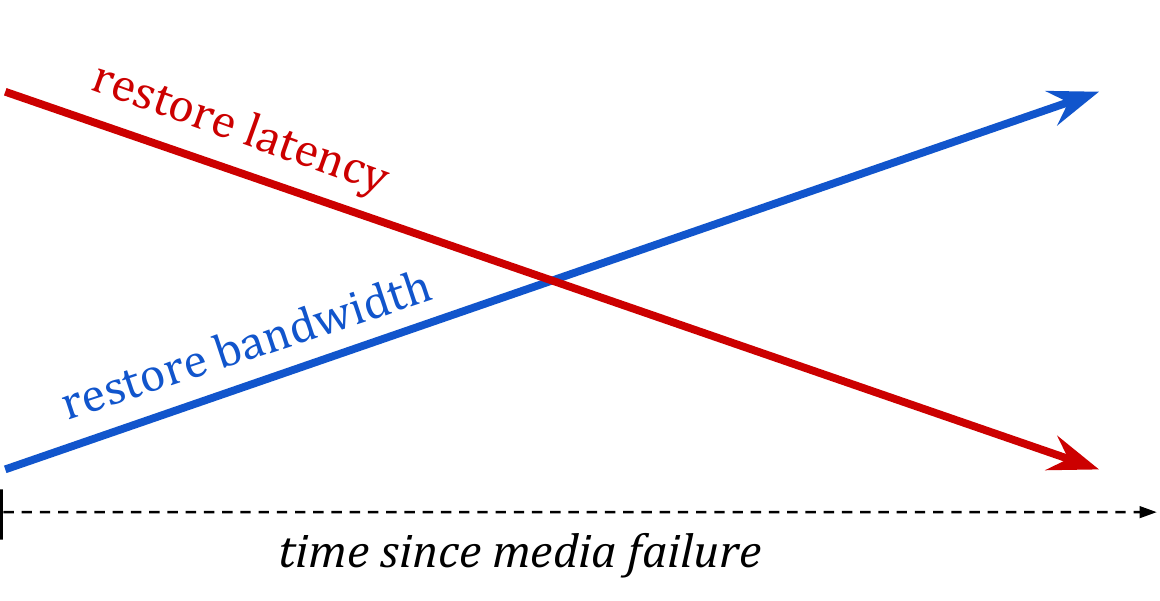}
  \caption{Restore behavior optimized for latency and bandwidth}
\label{fig:latency_bandwidth}
\vspace{-0.3cm}
\end{figure}

Restore latency is defined as the additional delay imposed on the page reads and writes of an individual transaction due to restore operations.
Hence, it follows that if a single page can be read and restored in the same time it takes to just read it, the restore latency is zero---this is the “gold standard” of restore performance and availability.
For a single transaction, restore latency can be reduced by setting a small segment size---e.g., a single page.
However, this is not the optimal behavior when considering average restore latency across all transactions.
Therefore, restore bandwidth, i.e., the number of bytes restored per second, must also be optimized.
The optimized restore behavior is illustrated in Fig.~\ref{fig:latency_bandwidth}: in the beginning of the restore process, pages which are needed more urgently should be restored first, so that restore latency is decreased; towards the end, less and less transactions must wait for restore, so the system can effectively increase restore bandwidth while a low restore latency is maintained.

It is also worth noting that devices with low latency and inherent support for parallelism, e.g., solid-state drives, make these trade-offs less pronounced.
This does not mean, however, that instant restore is any less significant for such devices---a point which we would like to emphasize with the next two paragraphs.

As discussed earlier, previous restore techniques suffered from two deficiencies: inefficient access pattern and lack of incremental and on-demand recovery.
Solid-state devices shorten the efficiency gap between restore algorithms with sequential and random access, but this gap will never be entirely closed---if anything, thanks to the locality and predictability of sequential access.

As for the second deficiency, low-latency devices directly contribute to the reduction of restore latency, because the time to recover a single segment is reduced with faster access to backup and log archive runs.
Therefore, with instant restore, any improvement on I/O latency directly translates into lower time to repair---as perceived by a single transaction---and thus higher availability.
Non-incremental techniques, where the restore latency is basically the time for complete recovery, do not benefit as much from low-latency storage hardware when it comes to improving restore latency.

In terms of latency and bandwidth trade-off in the instant restore algorithm, the first choice to be made is the segment size.
In order to simplify the tracking of restore progress with a simple bitmap data structure, a fixed segment size must be chosen when initializing the restore manager.
We recommend choosing a minimum size such that acceptable bandwidth is delivered even for purely random access, but not too many segments exist such that the bitmap would be too large; e.g., 1~MB seems like a reasonable choice in practice.

In order to exploit opportunities for increasing bandwidth, multiple contiguous segments should be restored in a single step when applicable.
One technique to achieve that dynamically and adaptively is to simply run single-pass restore concurrently with instant restore.
Since the two processes rely on the same algorithm, no additional code complexity is required.
Furthermore, the coordination between them is essentially the same as that between concurrent instant restore processes---they both rely on the buffer pool and the segment recovery bitmap.
Section~\ref{sec:Coordination} exposes details of that coordination.

Alternatively, the scheduler component of the restore manager can employ a preemptive policy, where multiple contiguous segments are restored as long as no requests arrive in its incoming queue.
As shown empirically in Section~\ref{sec:Experiments}, this simple technique automatically prioritizes latency in the beginning of the restoration process, when the most important pages are being requested; then, as less and less transactions access data not yet restored, bandwidth is increased gradually with larger restoration units.
This technique essentially delivers the behavior presented in Fig.~\ref{fig:latency_bandwidth}.

In terms of log archive access, the size of initial runs poses an important trade-off between minimizing merge effort and minimizing the lag between generating a log record and persisting it into the log archive.
In order to generate larger runs, log records must be kept longer in the in-memory sort workspace.
On the other hand, correct recovery requires that all log records up to the time of device failure be properly archived before restore can begin; thus, smaller initial runs imply lower restore latency for the first post-failure transactions.
While this choice is important, simple techniques largely mitigate these concerns.
One option is to enable access to log records while they are still in the main-memory sort workspace.
This is possible because, as discussed in Section~\ref{sec:FailureClasses}, a media failure does not incur loss of the server process and its in-memory contents.
Alternatively, single-page repair could be used to replay log records that are not yet archived when a segment is restored.
As with concurrent single-pass restore, these individual recovery techniques are orthogonal and can thus be applied concurrently with minimal coordination.
Using the techniques sketched above, the lag incurred by the archiving process would be minimized.

Besides these concerns specific to instant restore, established techniques to choose initial run size and merge fan-in based on device characteristics directly apply \cite{bib:GraefeQueryEvaluation}.
This is mainly because the access pattern of instant restore basically resembles that of an external sort followed by a merge join.

\subsection{Coordination of recovery actions}\label{sec:Coordination}
As mentioned briefly above, the segment recovery bitmap enables the coordination of concurrent restore processes, allowing configurable scheduling policies.
Another important aspect to be considered is the coordination among restore and the other recovery modes summarized in Table~\ref{tab:FailureClasses}.
This section discusses how to coordinate all such recovery actions without violating transactional consistency.

The first failure class---transaction failure---is the easiest to handle because its recovery is made transparent to the other classes thanks to rollback by compensation actions, as introduced in ARIES \cite{bib:Mohan92ARIES} and refined in the multi-level transaction model \cite{bib:WeikumVossen}.
The implication is that recovery for the other failure classes must distinguish only between uncommitted and committed transactions.
Transactions that abort are simply considered committed---it just happens that they revert all changes they made, i.e., they ``commit nothing''.
Therefore, for the purposes of instant restore, transactions that issue an abort behave exactly like any other in-flight transaction, including those that started after the failure: they hold locks to protect their reads and updates and access data through the buffer pool, which possibly triggers segment restoration as described in Section~\ref{sec:Restore}.

As for the other three classes, recovery coordination using the techniques presented in this work requires a distinction between two general forms of recovery: using a transaction log and using the indexed log archive.
The former is assumed to be a linear data structure ordered strictly by LSN and containing embedded chains among log records of the same transaction and of the same page---whether it resides on active or archive devices does not matter for this discussion.
Single-page repair and restart after a system failure both use the transaction log, and whether the old page image comes from a backup or from the persistent database also does not matter for this discussion.
Since they perform log replay on a single page at a time, they are coordinated using the fix and unfix operations of the buffer pool.
Because replaying updates on a page requires an exclusive latch, the same page cannot be recovered concurrently by different recovery processes of any kind.
Furthermore, tracking the page LSN of the fixed buffer pool frame guarantees that updates are never replayed more than once and that no updates are missed.
This mechanism ensures correctness of concurrent restart and single-page repair processes.


The second form of recovery---using the indexed log archive---is used solely for instant restore at the segment granularity.
Here, a segment, whose size is fixed when a failure is detected, is the unit of recovery, and coordination relies on the segment recovery bitmap.
Using two states---restored and not restored---avoids restoring a segment more than once in sequence, but additional measures are required to prevent that from happening concurrently.
One option is to simply employ a map with three states, the additional one being simply ``undergoing restore''.
A thread encountering the ``not restored'' state attempts to atomically change it to ``undergoing restore'': if it succeeds, it initiates the restore request for the segment in question; otherwise, it simply waits until the state changes to ``restored''.

Alternatively, coordination of segment restore requests can reuse the lock manager.
A shared lock is acquired before verifying the bitmap state, and, in order to restore a segment, the shared lock must be upgraded to exclusive with an unconditional request.
The thread that is granted the upgrade is then in charge of restoration, while the others will automatically wait and be awoken by the lock protocol, after which they see the ``restored'' state.

While the segment recovery bitmap provides coordination of concurrent restore processes, the buffer fix protocol is again used to coordinate restore with the other recovery modes.
Concomitant restart and restore processes may occur in practice because some failures tend to cause related failures.
A hardware fault, for instance, may not only corrupt persistent data but also cause an operating system crash.
In this case, the recovery processes will be automatically coordinated with the methods described above.
Restart recovery will fix pages in the buffer pool prior to performing any redo or undo action.
The fix call, in turn, will issue a read request on the device.
If the device has failed, the restore manager will intercept this request and follow the restore protocol described above.
Only after the containing segment is restored, the fix call returns.
After that, the page may still require log replay in the redo phase of restart, which is fine---the two recovery modes will simply replay different ranges of the page's history.

\subsection{Summary of instant restore}
Instant restore is enabled by an indexed log archive data structure that can be generated online with very low overhead.
By partitioning data pages into segments, the recovery algorithm provides incremental and on-demand access to restored data.
The algorithm requires a simple bitmap data structure to keep track of progress and coordinate restoration of individual segments under configurable scheduling policies.

The generalized nature of instant restore enables a wide range of choices for trading restore latency and bandwidth.
These choices can be made adaptively and robustly by the system using simple techniques.
Moreover, while instant restore mitigates many of the issues with high-capacity hard disks, making them a more attractive option, it still benefits greatly from modern storage devices such as solid-state drives.
Therefore, the technique is equally relevant for improving availability with any kind of storage hardware.

Lastly, the restore processes can be easily coordinated with processes from other recovery modes---the independence of these modes and the integrated coordination using the buffer pool ensure transaction consistency in the presence of an arbitrary mix of failure classes.

\section{Experiments}\label{sec:Experiments}
Our experimental evaluation covers three main measures of interest during recovery from a media failure: restore latency, restore bandwidth, and transaction throughput.
Moreover, we evaluate the overhead of log archiving with sorting and indexing in order to assess the cost of instant restore during normal processing.
Before presenting the empirical analysis, a brief summary of our experimental environment is provided.

\subsection{Environment}
We implemented instant restore in a fork of the Shore-MT storage manager \cite{bib:Johnson09ShoreMT} called \emph{Zero}.
The code is available as open source \footnote{\url{http://github.com/caetanosauer/zero}}.
The workload consists of the TPC-C benchmark as implemented in Shore-MT, but adapted to use the Foster B-tree~\cite{bib:FosterBtree} data structure for both table and index data.

All experiments were performed on dual six-core CPUs with two thread contexts each, yielding support for 24 hardware threads.
The system has 100~GB of high-speed RAM and several Samsung 840 Pro 250~GB SSDs connected to a dedicated I/O controller. 
The operating system is Ubuntu Linux 14.04 with Kernel 3.13.0-68 and all code is compiled with \texttt{gcc} 4.8 and \texttt{-O3} optimization.

The experiments all use the same workload, with media failure and recovery set up as follows.
Initial database size is 100~GB, with full backup and log archive of the same size---i.e., recovery starts from a full backup of 100~GB and must replay roughly the same amount of log records.
Log archive runs are a little over 1.5~GB in size, resulting in 64 inputs in the restore merge logic.
All persistent data is stored on SSDs and 24 hardware threads are used at all times.
The benchmark starts with a warmed-up buffer pool, whose total size we vary in the experiments.

\subsection{Restore latency}

Our first experiment evaluates restore latency, as defined in Section~\ref{sec:Restore}, by analyzing the total latency of individual transactions before and after a media failure.
The hypothesis under test is that average transaction latency immediately following a media failure is in the order of a few seconds or less, after which is gradually decreases to the pre-failure latency.
Furthermore, with larger memory, i.e., where a larger portion of the working set fits in the buffer pool, average latency should remain at the pre-failure level throughout the recovery process.

\begin{figure}
  \centering
  \includegraphics[width=\linewidth]{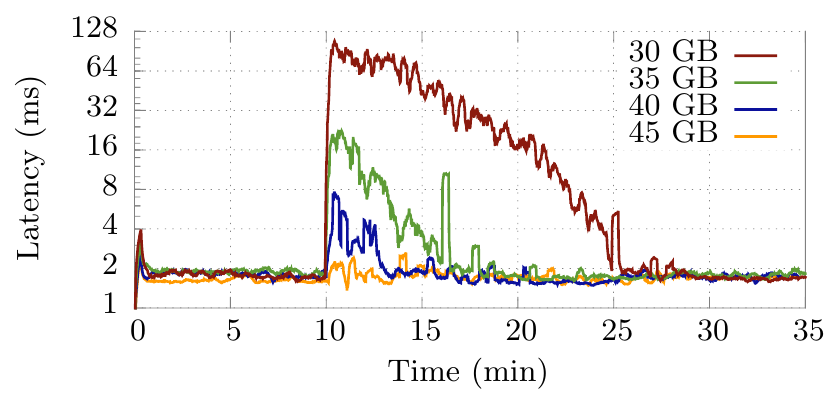}
  \caption{Transaction latency observed during instant restore: time series of average values.\label{fig:latency}}
  \vspace{-0.3cm}
\end{figure}

The results are shown in Fig.~\ref{fig:latency}.
After ten minutes of normal processing, during which the average latency is 1--2~ms, a media failure occurs.
The immediate effect is that average transaction latency spikes up (to about 100~ms in the buffer pool size of 30~GB) but then decreases linearly until pre-failure latency is reestablished.
For the largest buffer pool size of 45~GB, there is a small perturbation in the observed latency, but the average value seems to remain between 1 and 2~ms.
From this, we can conclude that for any buffer pool size above 45~GB, a media failure goes completely unnoticed.

\begin{figure}
  \centering
  \includegraphics[width=\linewidth]{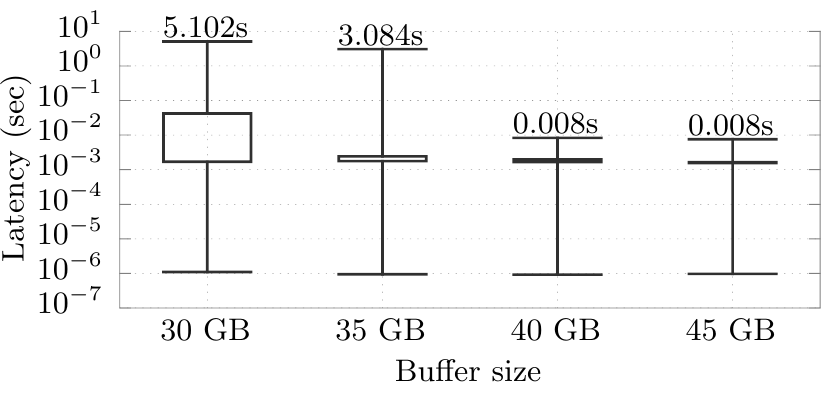}
  \caption{Transaction latency observed during instant restore: distribution of individual values.\label{fig:latency_distr}}
  \vspace{-0.3cm}
\end{figure}

\begin{figure*}[h]
  \begin{subfigure}[b]{0.5\textwidth}
    \centering
    \includegraphics[width=\linewidth]{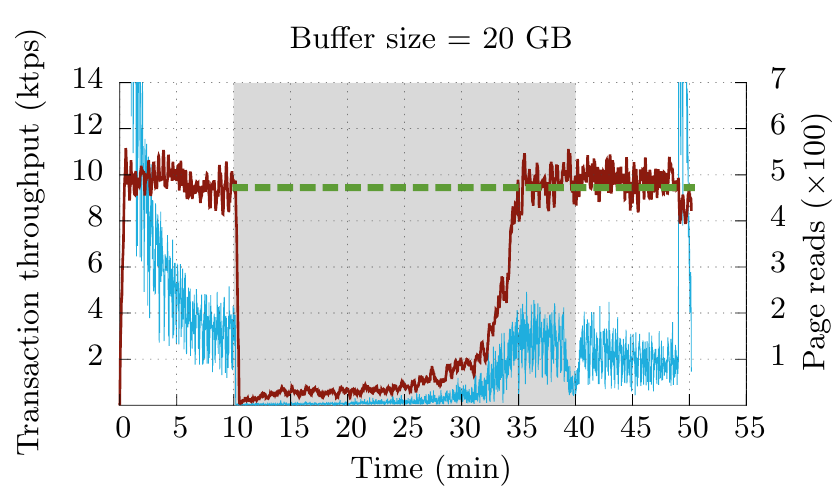}
  \end{subfigure}
  ~
  \begin{subfigure}[b]{0.5\textwidth}
    \centering
    \includegraphics[width=\linewidth]{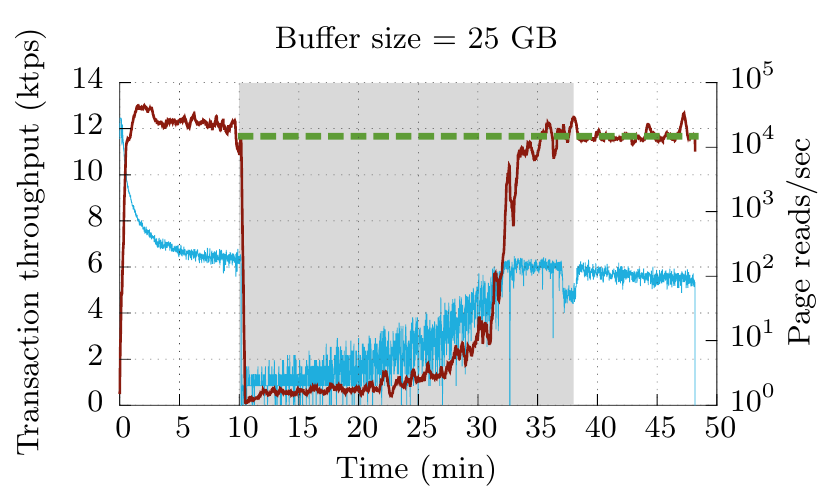}
  \end{subfigure}
  \\
  \begin{subfigure}[b]{0.5\textwidth}
    \centering
    \includegraphics[width=\linewidth]{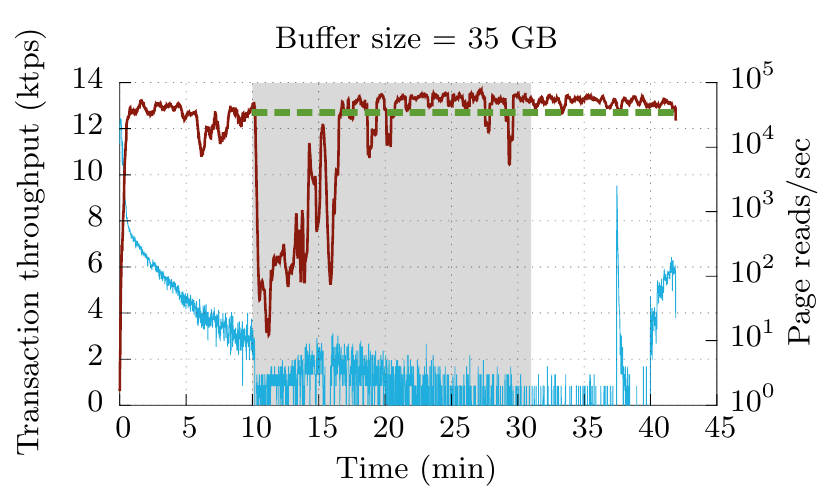}
  \end{subfigure}
  ~
  \begin{subfigure}[b]{0.5\textwidth}
    \centering
    \includegraphics[width=\linewidth]{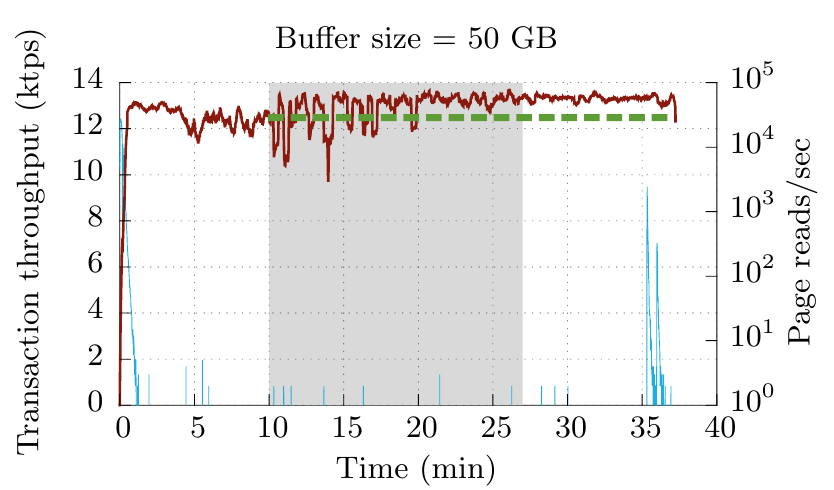}
  \end{subfigure}
  \\
  \begin{center}
    \vspace{-0.3cm}
    \includegraphics{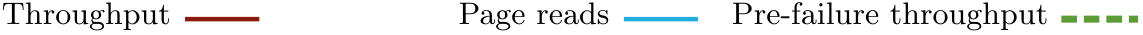}
  \end{center}
  \caption{Online restore and its impact on transaction throughput at different buffer pool sizes.}
  \label{fig:online_tput}
  \vspace{-0.3cm}
\end{figure*}

For this experiment, we also look at the distribution of individual latencies, in order to analyze the worst-case behavior.
As the plot on Fig.~\ref{fig:latency_distr} shows, the largest latency observed by a single transaction, with the smallest buffer pool, is 5.1~s.
The total recovery time, which is shown later in Fig.~\ref{fig:online_tput}, is in the range of 17--25 minutes---this is the restore latency incurred by single-pass restore.

These results successfully confirm our hypothesis: restore latency is reduced from 25 minutes to 5 seconds in the worst case, which corresponds to two orders of magnitude or two additional 9's of availability.
For the average case, still considering the smallest buffer pool, another order of magnitude is gained with latency dropping below 100~ms.

Note that the average restore latency is independent of total device capacity, and thus of total recovery time.
Therefore, the availability improvement could be in the order of four or five orders of magnitude in certain cases.
This would be expected, for instance, for very large databases (in the order of terabytes) stored on relatively low-latency devices.
In these cases, the gap between a full sequential read and a single random read---hence, between mean time to repair with single-pass restore and with instant restore---is very pronounced.

\subsection{Restore bandwidth}

Next, we evaluate restore bandwidth for the same experiment described earlier for restore latency.
The hypothesis here is that, in general, restore bandwidth gradually increases throughout the recovery process until it reaches the bandwidth of single-pass restore.
From these two general behaviors, two special cases are, again, the small and large buffer pools.
In the former, bandwidth may not reach single-pass speeds due to prioritization of low latency for the many incoming requests (recall that each buffer pool miss incurs a read on the replacement device, which, in turn incurs a restore request).
In the latter case, restore bandwidth should be as large as single-pass restore.


\begin{figure}
  \centering
  \includegraphics[width=\linewidth]{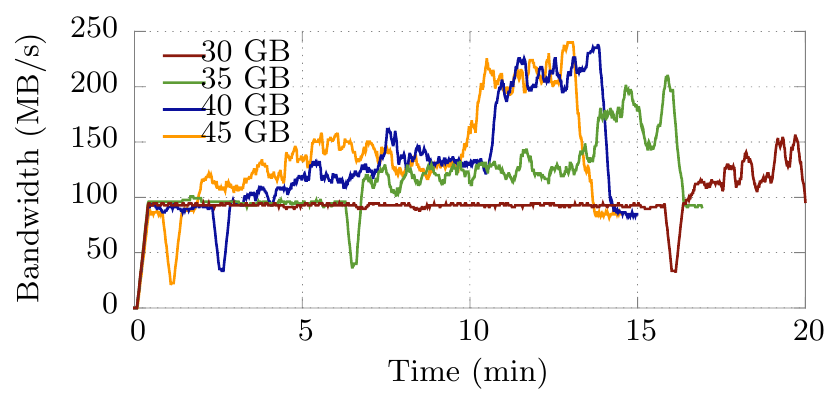}
  \caption{Bandwidth observed through the restore process.\label{fig:bandwidth_lines}}
  \vspace{-0.2cm}
\end{figure}

Fig.~\ref{fig:bandwidth_lines} shows the results of this experiment for four buffer pool sizes.
For the smallest buffer pool of 30~GB, restore bandwidth remains roughly constant in the first 15 minutes.
This indicates that during this initial period, most segments are restored individually in response to an on-demand request resulting from a buffer pool miss.
As the buffer size increases, the rate of on-demand requests decreases as restore progresses, resulting in more opportunities for multiple segments being restored at once.
In all cases, as predicted in the diagram of Fig.~\ref{fig:latency_bandwidth}, restore bandwidth gradually increases throughout the recovery process, reaching the maximum speed of 240~MB/s towards the end in the larger buffer pool sizes.

\subsection{Transaction throughput}

The next experiments evaluate how media failure and recovery impact transaction throughput with instant restore.
We take the same experiment performed in the previous sections and look at transaction throughput for each buffer pool size individually.
As instant restore progresses, transactions continue to access data in the buffer pool, triggering restore requests for each page miss.
Therefore, we expect that the larger the buffer pool is (i.e., more of the working set fits into main memory), the less impact a media failure has on transaction throughput.
This effect was already presented in the diagram of Fig,~\ref{fig:tput_simple}---the present section analyzes that in more detail.

Fig.~\ref{fig:online_tput} presents the results.
In the four plots shown, transaction throughput is measured with the red line on the left y-axis.
At minute 10, a media failure occurs, after which a green straight line shows the pre-failure average throughput.
The number of page reads per second is shown with the blue line on the right y-axis.
Moreover, total recovery time, which also varies depending on the buffer pool size, is also shown as the shaded interval on the x-axis.

The goal of instant restore in this experiment is to re-establish the pre-failure transaction throughput (i.e., the dotted green line) as soon as possible.
Similar to the evaluation on previous experiments, our hypothesis is that this occurs sooner the larger the buffer pool is.

The results show that for a small buffer pool of 20~GB, transaction throughput drops substantially, and it only regains the pre-failure level at the very end of the recovery process.
As the buffer size is increased to 25 and then 35~GB, pre-failure throughput is re-established at around minute 7, i.e., 1/3 of the total recovery time.
Lastly, for the largest buffer pool of 50~GB, the media failure does not produce any noticeable slowdown, as predicted in our hypothesis..

\subsection{Log archiving overhead}
\begin{figure}
  \includegraphics[width=\linewidth]{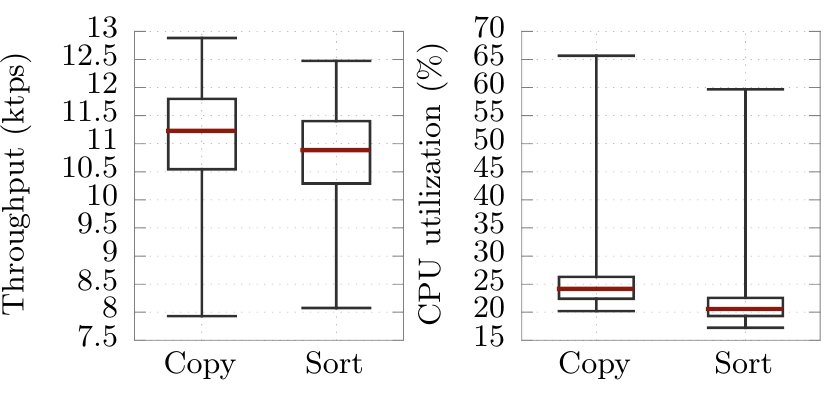}
  \caption{Overhead of log archiving}\label{fig:la_overhead}
  \vspace{-0.5cm}
\end{figure}

The overhead imposed on running transactions by sorting and indexing log records in the log archiving procedure is measured in the experiment of Figure \ref{fig:la_overhead}.
The chart on the left side shows average transaction throughput of a TPC-C execution using two variants of log archiving: with sorting and indexing, as required by instant restore, vs.~simply copying files.
The chart on the right side measures CPU utilization.
The charts show distributions of values using a candlestick chart; the box in the middle covers data points between the first and third quartiles of the distribution (i.e., half of the observations), while the extremities show the minimum a maximum values; the line in the middle of the box shows the median value.

As the chart on the left side shows, there is a small difference between a simplified implementation of traditional log archiving (i.e., plain filesystem copy) and log archiving with sorting and indexing.
The difference between the median points is less than 3\% (i.e., 11.2 vs.~10.9~ktps).
The CPU utilization measurement shows that it is proportional to transaction throughput, leading to the conclusion that the archiving process does not consume too much CPU.
Furthermore, note that the ``copy'' variant is quite primitive, since an industrial-strength implementation would incur additional overhead by compressing log records.
In that case, the overhead of our technique could be even less than 3\%.

\subsection{Summary of experiments}
We have shown that instant restore greatly improves upon the baseline single-pass restore algorithm.
Restore latency, which is the additional latency incurred on transactions by media recovery actions, is cut down by multiple orders of magnitude, which directly translates into the same improvement on availability.
Restore bandwidth adaptively and gradually approximates the maximum sequential speed as the recovery process progresses.
The same gradual improvement is observed for transaction throughput during media recovery.
These measures are equally affected by an increase in buffer pool size, up to a point where media failures cause no disruption at all on running transactions.
Lastly, we have shown that the online archiving procedure required by instant restore induces very little overhead (3\% or less) on normal processing---a small price to pay for the vast improvement in availability.

\section{Conclusions}\label{sec:Conclusion}
Instant restore improves perceived mean time to repair and thus database availability in the presence of media failures.
We identified two main deficiencies with traditional recovery techniques, such as the ARIES design \cite{bib:Mohan92ARIES}: (i) media recovery is very inefficient due to its random access pattern on database pages, which means that time to repair is unacceptably long; and (ii) data on a failed device cannot be accessed before recovery is completed.
The first deficiency was addressed in our previous work on single-pass restore \cite{bib:SinglePassRestore}, which introduces a partial sort order on the log archive, converting the random access pattern of log replay into sequential.

The second deficiency is addressed with the instant restore technique, which was first described in earlier work \cite{bib:InstantRecoverySynthesis} and discussed in more detail, implemented, and evaluated in this paper.
By generalizing single-pass restore and other recovery methods such as single-page repair, instant restore is the first media recovery method to effectively eliminate the two deficiencies discussed.
In comparison with traditional ARIES media restore, instant restore delivers not only the benefits of single-pass restore (i.e., substantially higher bandwidth and therefore shorter recovery time), but also much quicker access (e.g., seconds instead of hours) to the application working set after a failure.

Instant restore introduces a new organization of the log archive data structure, where log records are partially sorted and indexed.
Maintenance of this data structure incurs very little overhead and is performed continuously and online.

Our empirical analysis shows that instant restore is able to effectively deliver the efficiency of single-pass restore while cutting down restore latency by multiple orders of magnitude.
The experiments also analyze the impact of a failure on transaction throughput, which largely depends on the size of the working set in relation to the buffer pool size.
The results confirm our expectation that the pre-failure transaction throughput is re-established earlier as memory size increases---up to a point where a media failure goes completely unnoticed.
The net effect is that availability is greatly improved and the number of missed transactions due to media failures is significantly reduced.

\bibliographystyle{IEEEtran}
\bibliography{recovery}

\balance

\end{document}